\documentclass{ws-jnmp}

\usepackage{url}

\begin{document}

%\arttype{Letter} % default 'Article'

\markboth{Marco Frasca}{Exact solutions of classical scalar field equations}

\catchline{1}{1}{2009}{}{}
%
%%%%%%%%%%%%%%%%%%%%%%%%%%%%%%%%%%%%%%%%%%%%%%%%%%%%%%%%%%%%%%%%%%%%
\copyrightauthor{M. Frasca}

\def\bib{B\kern-.05em{I}\kern-.025em{B}\kern-.08em}
\def\btex{B\kern-.05em{I}\kern-.025em{B}\kern-.08em\TeX}

\title{Exact solutions of classical scalar field equations}

\markboth{WSPC}{Using World Scientific's \btex\/ Style File}

\author{Marco Frasca}

\address{Via Erasmo Gattamelata, 3 \\
00176 Roma (Italy)\\
\email{marcofrasca@mclink.it}}

\maketitle

\begin{history}
\received{(Day Month Year)}
\revised{(Day Month Year)}
\accepted{(Day Month Year)}
\end{history}

\begin{abstract}
We give a class of exact solutions of quartic scalar field theories. These solutions prove to be interesting as are characterized by the production of mass contributions arising from the nonlinear terms while maintaining a wave-like behavior. So, a quartic massless equation has a nonlinear wave solution with a dispersion relation of a massive wave and a quartic scalar theory gets its mass term renormalized in the dispersion relation through a term depending on the coupling and an integration constant. When spontaneous breaking of symmetry is considered, such wave-like solutions show how a mass term with the wrong sign and the nonlinearity give rise to a proper dispersion relation. These latter solutions do not change the sign maintaining the property of the selected value of the equilibrium state. Then, we use these solutions to obtain a quantum field theory for the case of a quartic massless field. We get the propagator from a first order correction showing that is consistent in the limit of a very large coupling. The spectrum of a massless quartic scalar field theory is then provided. From this we can conclude that, for an infinite countable number of exact classical solutions, there exist an infinite number of equivalent quantum field theories that are trivial in the limit of the coupling going to infinity.\end{abstract}


\section{Introduction}

Nonlinear partial differential equations enter into the description of so many physical effects to represent a vast subject for study. Indeed, not much exact solutions are known of such equations and most of the analysis are carried out using small perturbation theory both for classical and quantum solutions. But knowing exact solutions may help to understand the behavior of physical phenomena in a range of parameters where small perturbation theory just fails to produce significant results. This is a point generally emphasized when we work on question like QCD or just, on a more general side, how massive contribution could be produced beyond Higgs phenomenon.

So, our aims in this paper will be to show how an answer can be given to the question of mass simply for the effect of the nonlinearities and how such solutions can give rise to a significant quantum field theory in the limit of a strong coupling. This will not imply that the solutions we will present here should be the ones chosen by Nature but their very existence can be used as a track for further understanding and providing a serious means for other deep analysis.

A quartic field theory has a lot of applications ranging from condensed matter to quantum field theory \cite{zj}. It has got a first relevant application in the area of phase transitions \cite{lau1,lau2} and finally entered into particle physics through the pioneer works of Yoichiro Nambu and Jeffrey Goldstone \cite{nam,gold}. So, they have such a relevance that knowing some exact solutions may pave the way to a deeper understanding of the physical phenomena they enter.

Our current ideas about mass terms in quantum field theory take the move from a phase transition. A breaking of a symmetry selects the proper vacuum of a self-interacting quartic field and being this vacuum value not zero, an interaction with this field can give a mass term. But this could not be the only way a field can get a mass. Self-interaction could be another one as we will show. Besides, this property could be maintained at a quantum level provided a strong coupling in the self-interaction is considered. This will also produce a discrete spectrum of excitations. Indeed, we will get the Green function from a first order correction and the corresponding spectrum, at least for a quartic massless scalar field. In this way we will be able to reconnect such exact solutions to our recent proposal for a strong coupling analysis of a quantum field theory \cite{fra1}.

The main results of the paper is that, in the limit of the coupling going to infinity, there are an infinite number of equivalent quantum field theories, built from the corresponding subset of infinite exact classical solutions, that are trivial. Indeed, we get a Gaussian functional with the spectrum of a harmonic oscillator.

The paper is so structured. In sec.\ref{sec:classical} we give the relevant exact solutions having the property to contributing to the mass in the dispersion relations, due to the self-interaction term. In sec. \ref{sec:qft} we give the corresponding quantum field theory showing how this can be built starting from an infinite countable subset of the full set of classical solutions producing a trivial theory. Finally, our conclusions are given in sec. \ref{sec:conc}.

\section{Classical scalar field theories}
\label{sec:classical}

In the following we present wave-like exact solutions with peculiar dispersion relations. Mass terms or contributions to mass depend on the coupling in the nonlinear terms and so we can understand how self-interaction is truly effective in turning massless excitations into massive ones. We note that these solutions hold in any dimensionality provided that the integration constant has the proper dimension and the coupling has been made dimensionless.

\subsection{Massive and massless scalar field theories}
\label{sec:mass}

Let us consider the equation
\begin{equation}
-\Box\phi +\mu_0^2\phi+\lambda\phi^3 = 0
\end{equation}
being $\Box=-\partial_t^2+\Delta$ the wave operator, $\mu_0$ the mass of the field and $\lambda$ the coupling. By direct substitution one can verify that a solution is given by
\begin{equation}
\phi(x) = \pm\sqrt{\frac{2\mu^4}{\mu_0^2 + \sqrt{\mu_0^4 + 2\lambda\mu^4}}}{\rm sn}\left(p\cdot x+\theta,\sqrt{\frac{-\mu_0^2 + \sqrt{\mu_0^4 + 2\lambda\mu^4}}{-\mu_0^2 - 
   \sqrt{\mu_0^4 + 2\lambda\mu^4}}}\right)
\end{equation}
provided that
\begin{equation}
   p^2=\mu_0^2+\frac{\lambda\mu^4}{\mu_0^2+\sqrt{\mu_0^4+2\lambda\mu^4}}
\end{equation}
being $\theta$ and $\mu$ two integration constants and ${\rm sn}$ a Jacobi elliptical function that is periodic representing a nonlinear wave-like solution. From the dispersion relation we can see that the mass of the field gets a contribution from the coupling $\lambda$ and the integration constant $\mu$, so it arises from the self-interacting term as it becomes zero when $\lambda$ is taken to be zero (no interaction case).

These formulas are simplified setting the mass of the field to zero. As they will be used in the following, we give them here. The solution becomes
\begin{equation}
\label{eq:phis}
   \phi(x) = \pm\mu\left(\frac{2}{\lambda}\right)^\frac{1}{4}{\rm sn}(p\cdot x+\theta,i) 
\end{equation}
provided
\begin{equation}
   p^2=\mu^2\left(\frac{\lambda}{2}\right)^\frac{1}{2}.
\end{equation}
This solution shows how a massless field can become massive just from the self-interacting term. Already at the classical level, we get an arbitrary integration constant having the dimension of a mass.

\subsection{Spontaneous breaking of symmetry}
\label{sec:breaking}

We get spontaneous breaking of symmetry when the mass term is taken with a wrong sign to have
\begin{equation}
   -\Box\phi -\mu_0^2\phi +\lambda\phi^3= 0.
\end{equation}
This equation has the following exact wave-like solution
\begin{equation}
   \phi(x) =\pm v\cdot {\rm dn}(p\cdot x+\theta,i)
\end{equation}
provided that
\begin{equation}
   p^2=\frac{\lambda v^2}{2}
\end{equation}
being $v=\sqrt{\frac{2\mu_0^2}{3\lambda}}$. We see that the dispersion relation has the mass term with the right sign and we are describing oscillations around one of the selected solutions $\phi=\pm \sqrt{\frac{3}{2}}v$ as the Jacobi function ${\rm dn}$ is never zero.

\subsection{Fourier expansion of solutions}
\label{sec:fourier}

These solutions have a Fourier expansion being periodic functions. These expansions are widely known, being those of Jacobi elliptical functions. In this way one can present them as a superposition of plane waves. So, one has \cite{gra}
\begin{equation}
   \phi(x) = \frac{2\pi}{kK(k)}\sum_{n=0}^\infty\frac{q^{n+\frac{1}{2}}}{1-q^{2n+1}}
   \sin\left[(2n+1)\frac{\pi}{2K(k)}p\cdot x\right]
\end{equation}
being $q=e^{-\pi\frac{K'(k)}{K(k)}}$, 
$k=\sqrt{\frac{-\mu_0^2 + \sqrt{\mu_0^4 + 2\lambda\mu^4}}{-\mu_0^2 - \sqrt{\mu_0^4 + 2\lambda\mu^4}}}$, $K'(k)=K(\sqrt{1-k^2})$ and finally $K(k)=\int_0^{\pi/2}dx/\sqrt{1-k^2\sin x}$. The reason why such a Fourier series is interesting is that, in the rest reference frame, taking ${\bf p}=0$, one has
\begin{equation}
   \phi(t,0) = \frac{2\pi}{kK(k)}\sum_{n=0}^\infty\frac{q^{n+\frac{1}{2}}}{1-q^{2n+1}}
   \sin\left[(2n+1)\frac{\pi}{2K(k)}mt\right]
\end{equation} 
having set
\begin{equation}
   m=m(\mu_0,\mu,\lambda)=\sqrt{\mu_0^2+\frac{\lambda\mu^4}{\mu_0^2+\sqrt{\mu_0^4+2\lambda\mu^4}}}
\end{equation}
that is the ``renormalized mass'' for these classical field excitations. From this expansion one can read out a kind of mass spectrum
\begin{equation}
   \epsilon_n=(2n+1)\frac{\pi}{2K(k)}m.
\end{equation}
In order to interpret this as a true mass spectrum we need a quantum field theory. In the next section we will show that this is indeed the case and these solutions represent essentially free massive particles. These equations simplify significantly taking $\mu_0=0$. As we will use these formulas in the following, we will give them here. One has
\begin{equation}
   \phi(x)=\pm\mu\left(\frac{2}{\lambda}\right)^\frac{1}{4}\sum_{n=0}^\infty(-1)^n\frac{2\pi}{K(i)}
   \frac{e^{\left(n+\frac{1}{2}\right)\pi}}
   {1+e^{-(2n+1)\pi}}\sin\left((2n+1)\frac{\pi}{2K(i)}p\cdot x\right)
\end{equation}
so that
\begin{equation}
\label{eq:ser}
   \phi(t,0)=\pm\mu\left(\frac{2}{\lambda}\right)^\frac{1}{4}\sum_{n=0}^\infty(-1)^n\frac{2\pi}{K(i)}
   \frac{e^{\left(n+\frac{1}{2}\right)\pi}}
   {1+e^{-(2n+1)\pi}}\sin\left((2n+1)\frac{\pi}{2K(i)}\left(\frac{\lambda}{2}\right)^\frac{1}{4}\mu t\right)
\end{equation}
and a ``mass spectrum''
\begin{equation}
   \epsilon_n = (2n+1)\frac{\pi}{2K(i)}\left(\frac{\lambda}{2}\right)^\frac{1}{4}\mu.
\end{equation}
Finally, we consider the case for spontaneous breaking of symmetry having
\begin{equation}
   \phi(x)=\frac{\pi}{2K(i)}+\frac{2\pi}{K(i)}\sum_{n=1}^\infty\frac{(-1)^2e^{-n\pi}}{1+e^{-2n\pi}}
   \cos\left(2n\frac{\pi}{2K(i)}p\cdot x\right)
\end{equation}
and so
\begin{equation}
   \phi(t,0)=\frac{\pi}{2K(i)}+\frac{2\pi}{K(i)}\sum_{n=1}^\infty\frac{(-1)^2e^{-n\pi}}{1+e^{-2n\pi}}
   \cos\left(2n\frac{\pi}{2K(i)}\frac{\mu_0}{\sqrt{3}} t\right)
\end{equation}
giving rise to a ``mass spectrum'' \cite{fra2}
\begin{equation}
   \epsilon_n=n\frac{\pi}{K(i)}\frac{\mu_0}{\sqrt{3}}
\end{equation}
with $n=0,1,2,\ldots$.

\section{Quantum field theory of a massless scalar field theory}
\label{sec:qft}

In this section we analyze the simplest case of a massless quartic scalar field and we exploit the corresponding quantum field theory. The idea is to do a series development around the exact solution given above and evaluate the corresponding first order correction to see how this term will go to depend on the coupling $\lambda$. This will permit us to obtain a physical understanding of such an expansion. We note that the coupling $\lambda$ is dimensionless only for D=4. So, in the following we take the product $\lambda^\frac{1}{4}\mu$ having always the dimension of a mass.

We consider the generating functional
\begin{equation}
    Z[j] = N\int[d\phi] e^{i\int d^Dx
    \left[\frac{1}{2}(\partial\phi)^2-\frac{\lambda}{4}\phi^4+j\phi\right]}
\end{equation}
being $N$ a normalization constant, and we take the substitution $\phi=\phi_c+\delta\phi+O(\delta\phi^2)$ being $\phi_c$ the classical solution given in eq.(\ref{eq:phis}). We will recover the results given in \cite{fra1} and the first higher correction as it should be. After the substitution has done one has
\begin{equation}
\label{eq:zf}
    Z[j]=e^{i[S_c+\int d^Dx j\phi_c]} \int[d\delta\phi] e^{i\int d^Dx
    \left[\frac{1}{2}(\partial\delta\phi)^2
    -\frac{3}{2}\phi_c^2(\delta\phi)^2+j\delta\phi\right]}+O((\delta\phi)^3).
\end{equation}
So, we can see that we can accomplish a fully integration setting
\begin{equation}
    \delta\phi=\delta\phi_0+\int d^Dy\Delta_1(x-y)j(y),
\end{equation}
in the path integral, being
\begin{equation}
\label{eq:d1}
    -\Box\Delta_1(x-y)+3\lambda\phi_c^2(x)\Delta_1(x-y)=\delta^D(x-y).
\end{equation}
This equation can be solved exactly writing down
\begin{equation}
   G_n(x)=-\delta^{D-1}(x)\theta(t)\frac{1}{\mu(2^3\lambda)^{\frac{1}{4}}}
   {\rm cn}\left[\left(\frac{\lambda}{2}\right)^{\frac{1}{4}}\mu t+(4n+1)K(i),i\right]
   {\rm dn}\left[\left(\frac{\lambda}{2}\right)^{\frac{1}{4}}\mu t+(4n+1)K(i),i\right]
\end{equation}
when the phase of the exact solution is taken to be $\theta_n=(4n+1)K(i)$. This identifies an infinite class of quantum field theories for which we are able to compute starting from a subset of infinite countable exact solutions of the classical theory. All such theories are equivalent. So, in the end we have
\begin{equation}
   Z_n[j]=Z_n[0]e^{i[\int d^Dx j(x)\phi_c^{(n)}(x)+\int d^Dxd^Dyj(x)G_n(x-y)j(y)]}+O((\delta\phi)^3)
\end{equation}
that has the Gaussian form of a free theory. In order to compute the spectrum of the theory, we note that one can write
\begin{equation}
   G_n(x)=-\delta^{D-1}(x)\theta(t)\frac{1}{2\mu^2}\left.
   \frac{d}{du}\phi^{(n)}_c(u,0)\right|_{u=\left(\frac{\lambda}{2}\right)^{\frac{1}{4}}\mu t}. 
\end{equation}
Using eq.(\ref{eq:ser}) one has finally
\begin{equation}
   G_n(x)=-\delta^{D-1}(x)\theta(t)\frac{1}{2\mu}
   \left(\frac{2}{\lambda}\right)^\frac{1}{4}\frac{\pi^2}{K(i)^2}\sum_{n=0}^\infty(-1)^n(2n+1)
   \frac{e^{\left(n+\frac{1}{2}\right)\pi}}{1+e^{-(2n+1)\pi}}
   \cos\left((2n+1)\frac{\pi}{2K(i)}\left(\frac{\lambda}{2}\right)^\frac{1}{4}\mu t+\theta_n\right)
\end{equation}
that has the required form
\begin{equation}
   G_n(x)=-\delta^{D-1}(x)\theta(t)\sum_{n=0}^\infty B_n e^{-i\epsilon_n t} + c.c.
\end{equation}
with
\begin{equation}
   \epsilon_n=(2n+1)\frac{\pi}{2K(i)}\left(\frac{\lambda}{2}\right)^\frac{1}{4}\mu
\end{equation}
that can now be identified with a mass spectrum. We have a quantum theory of free massive particles. We note that we have recovered the same result obtained with a strong coupling expansion applied to quantum field theory \cite{fra1}. So, this theory is consistent in the limit of a very large coupling $\lambda$. We can state this conclusion in a different way: {\it An infinite class of quantum field theories can be built, from an infinite subset of exact solutions of a classical massless quartic field theory, that are trivial in the limit of the coupling going to infinity}.

%Inserted on 20-11-2010
Finally, we can prove that quantum fluctuations have a Goldstone mode. We work on the lines of Ref.\cite{zoli}. From the generating functional (\ref{eq:zf}) we can take for the fluctuations a series in eigenmodes
\begin{equation}
   \delta\phi(x)=\sum_n c_n\chi_n(x)
\end{equation}
so that, for each eigenmode, one has to solve the eigenvalue equation
\begin{equation}
   -\Box\chi_n(x)+3\lambda\phi_c^2(x)\chi_n(x)=\epsilon_n\chi_n(x).
\end{equation}
Already for the Green function above we have shown the existence of the solution for $n=0$. So, we conclude that a zero mode exists due to translational invariance. It would be interesting to know the fate of this mode when such a scalar field interacts with a gauge field \cite{fra3}.

One of the interesting consequences of this result is that a quantum field theory built with Jacobi functions is a free theory exactly as for the case of sine and cosine functions. But this result should be expected as the latter are a particular case of the former.

\section{Conclusions}
\label{sec:conc}

In this paper we have proved an interesting result, consistent with the behavior of scalar theories in the large coupling regime, that there exist an infinite class of equivalent quantum field theories, arising from exact solutions of the corresponding classical equation of motion, that are trivial. They have a Gaussian generating functional and the spectrum of a harmonic oscillator. But the result to be emphasized is that such free particles are massive when the theory we started with has no mass term. When a mass term is present, the theory show a correcting term to the mass itself that depends on the coupling and an integration constant and that goes to zero when such a coupling goes to zero. This means that mass terms arise from the self-interaction of the field and, as this becomes more and more important, with the failure of small perturbation theory, field excitations acquire mass. This in turn should imply that, if we want to go to higher orders, these excitations are the quantum states to start with.

\section*{Acknowledgments}

This paper was worked out after recollecting several results that I posted on Dispersive Wiki (\url{http://tosio.math.toronto.edu/wiki/index.php/Main_Page}) managed by Terry Tao and Jim Colliander. I take this chance to thank them for permitting me to post some of my results there.

% for BiBTeX users
%\bibliographystyle{ws-jnmp}   % link to ws-jnmp.bst
%\bibliography{pattern2}       % link to pattern2.bib

\end{document}